%% file: main.tex
\documentclass[10pt]{article}

\usepackage[margin=0.88in]{geometry}
\usepackage{microtype}
\usepackage{amsmath,amssymb}
\usepackage{booktabs}
\usepackage{array}
\usepackage{enumitem}
\usepackage{float}
\usepackage{xcolor}
\usepackage{graphicx}
\usepackage{tikz}
\usetikzlibrary{arrows.meta,calc,positioning,shapes.geometric}
\usepackage{listings}
\usepackage[numbers,sort&compress]{natbib}

\usepackage[colorlinks=true,allcolors=loopblue]{hyperref}
\usepackage[nameinlink,noabbrev]{cleveref}

\definecolor{loopblue}{HTML}{215A8E}
\definecolor{loopred}{HTML}{B64832}
\definecolor{loopgray}{HTML}{4B5563}
\definecolor{looplight}{HTML}{F4F6F8}

\hypersetup{
  pdftitle={Loop Engineering: Version-Bound, Evidence-Gated Coordination for Production Coding Agents},
  pdfauthor={Yupeng Pang; Meiling Chen; Yang Yang; Wei Fu},
  pdfsubject={Agentic software engineering and multi-agent coordination},
  pdfkeywords={Loop Engineering, coding agents, software engineering, shared state, evidence, human-in-the-loop}
}

\setlist{nosep,leftmargin=1.5em}
\title{\textbf{Loop Engineering: Version-Bound, Evidence-Gated Coordination\\for Production Coding Agents}}
\author{
  Yupeng Pang\thanks{Corresponding author: \texttt{bkidyy@gmail.com}}
  \quad Meiling Chen
  \quad Yang Yang
  \quad Wei Fu\\[0.45em]
  MewDesign
}
\date{July 2026}

\begin{document}
\maketitle

\begin{abstract}
Coding agents can edit repositories and execute tests, yet production software work extends
far beyond code generation. It requires durable task context, scoped authority, version-aware
approval, source-independent verification, interruption recovery, and accountable handoffs
between humans and automated workers. We present \emph{Loop Engineering}, a method for
operating multiple asynchronous agent loops over shared, auditable repository state without
relying on private conversational handoffs. Each loop is specified by six elements: a trigger,
a context projection, an authority boundary, a verifier, a stop condition, and a write-back
rule. Versioned procedural contracts, called Skills, encode reusable operating policy; a
Harness bounds the available action surface and collects evidence; human gates authorize
high-consequence transitions against exact artifact versions.

We report a design-and-experience study from MewDesign, a revenue-generating commercial AI
design product. Five specialized loops cover planning, scoped implementation, review, test
release, and reconciliation. A cross-sectional audit contains 50 durable marker records across
26 GitHub work items. Five pull requests reached a version-bound test lock: three proceeded to
test releases with runtime evidence from sources distinct from the deployment request, and two
stopped before merge. A detailed drift episode shows one observed fail-closed path: a new
commit invalidated prior approval and evidence, and the workflow refused to reuse stale
continuous-integration results. The study is not a controlled benchmark and does not establish
causal productivity or general safety. It provides early evidence of feasibility, traceability,
and bounded progression in one production setting.
\end{abstract}

\section{Introduction}

The visible capability of a coding agent is often measured by the patch it can produce.
Real software delivery, however, includes the less visible work around that patch: deciding
whether a task is ready, reconstructing the latest context, limiting scope, checking the exact
revision, obtaining authorization, following build and deployment systems, verifying runtime
state, and leaving enough evidence for another worker to continue later.

This is increasingly an operational rather than speculative problem. AIDev reports hundreds
of thousands of agent-authored pull requests across more than one hundred thousand public
repositories \citep{li2026aidev}. The resulting engineering question is no longer only whether
an agent can produce a plausible patch. It is whether organizations can delegate work while
retaining inspectable state, bounded authority, and accountable acceptance.

This distinction became operationally important while building
\href{https://mew.design/}{MewDesign}, a commercial AI design product. The development
workflow moved from IDE assistance, where a human manually
advanced every step, to command-line coding agents capable of long repository-level tasks.
Longer agent trajectories reduced the amount of direct editing but did not remove the human
as the workflow scheduler. The bottleneck moved from the \emph{inner loop} of code generation
to the \emph{outer loop} of state, authority, verification, and release.

Popular multi-agent designs often address this outer loop by assigning conversational roles:
a planner tells a coder what to do, a reviewer comments on the result, and a manager declares
completion \citep{qian2024chatdev,hong2024metagpt,wu2024autogen}. This is useful for many
tasks, but it creates a fragile coordination substrate for software operations. Role labels
do not guarantee independent judgment. Summaries between agents can lose source context.
Agents using the same model can amplify a shared false premise. Recent studies identify
coordination, commitment, and verification failures in multi-agent systems
\citep{cemri2025mast,khatua2026cooperbench}.

Loop Engineering takes a different systems stance:

\begin{quote}
\textbf{Agents do not rely on private conversational handoffs. They coordinate through typed,
durable records and independently inspectable evidence.}
\end{quote}

The shared substrate is the repository collaboration system already used by engineers:
issues, comments, labels, pull requests, commit identifiers, reviews, CI checks, deployment
artifacts, and runtime observations. Specialized loops wake on explicit state predicates,
reconstruct context from authoritative sources, act within bounded authority, write evidence
back to the shared state, and stop. They need not preserve a private conversation or trust
another agent's summary.

This paper is a design-and-experience report rather than a benchmark paper. It makes four
contributions:

\begin{enumerate}
  \item It specifies a six-part operational model for asynchronous agent loops, including
  exact-version authorization, evidence-gated transitions, and explicit invalidation.
  \item It presents a production architecture in which repository state is the coordination
  substrate, Skills encode versioned procedure, the Harness bounds capabilities, and humans
  retain high-consequence authority.
  \item It reports an artifact audit of five deployed loops, including three test releases,
  two deliberate stops, and an exact-head drift trace, with the strength of each empirical
  conclusion stated explicitly.
  \item It distinguishes contract-level, tool-mediated, and human-enforced controls and derives
  evaluation criteria beyond patch correctness, including authority, recoverability, evidence
  freshness, and time to verified outcome.
\end{enumerate}

We do not claim that one product deployment establishes causal productivity gains or general
superiority over other agent architectures. The contribution is a method, a concrete
realization, and early production evidence about properties that benchmark-only evaluations
rarely observe.

\section{Background and Problem}

\subsection{From Patch Generation to Workflow Progression}

SWE-bench made issue resolution in real repositories a standard evaluation target
\citep{jimenez2024swebench}. SWE-agent showed that the agent-computer interface materially
affects an agent's ability to navigate, edit, and test repositories \citep{yang2024sweagent};
OpenHands developed a general platform for software agents \citep{wang2025openhands}.
Agentless demonstrated that a simpler localization-repair-validation pipeline can outperform
more elaborate agents under some benchmark conditions \citep{xia2024agentless}. SWE-Lancer
further connects software tasks to real economic value \citep{miserendino2025swelancer}.

These efforts illuminate task competence, but production progression introduces additional
questions:

\begin{itemize}
  \item Which external state makes a task eligible now?
  \item Which plan version has actually been approved?
  \item Which repository revision was reviewed and authorized?
  \item What can this worker mutate, and what is expressly forbidden?
  \item Does a successful request mean the intended runtime state exists?
  \item Can an interrupted or repeated run resume without duplicating effects?
  \item What evidence lets a different worker, or a human, trust the handoff?
\end{itemize}

A system that solves code generation but leaves these questions implicit still requires a
human to drive every transition. Loop Engineering is aimed at this outer-loop problem.

\subsection{Failure Model}

We began with a practical failure model rather than a desired agent count. The following
failure classes recurred in long-running software work:

\paragraph{Context drift.}
The issue, plan, base branch, or pull-request head changes after an agent has reasoned about
it. A previously valid review or approval becomes stale.

\paragraph{Authority drift.}
An agent interprets a broad objective as permission to expand scope, modify another worker's
branch, touch production, or use an unrelated operational capability.

\paragraph{Self-attested completion.}
The worker that performed an action also declares success from an intermediate signal. For
example, an HTTP 200 from a deployment webhook proves request acceptance, not that a new
container digest is serving healthy traffic.

\paragraph{Duplicate effects.}
Schedulers, network calls, and agent tasks may be retried. Without leases, stable effect keys,
and reconcilable write-back, a resumed run can create duplicate pull requests, repeated
deployments, or conflicting notifications.

\paragraph{Correlated review.}
A ``reviewer'' with the same premise and context as the builder may restate the builder's
conclusion. Prompt-level role separation is not evidence-level independence.

\paragraph{Conversation loss.}
Private agent memory and inter-agent summaries are not durable engineering records. They
are difficult for humans to inspect and fragile under context truncation or process restart.

The system should fail closed under these conditions: uncertainty, drift, or missing evidence
must stop progression rather than broaden autonomy.

\section{The Loop Engineering Model}

\subsection{Shared State}

Let the authoritative engineering state at time $t$ be
\[
G_t = \langle I, P, B, H, R, Q, D, E \rangle_t,
\]
where $I$ is issue state, $P$ is the versioned plan, $B$ denotes branch relationships,
$H$ is the exact head revision, $R$ is review state, $Q$ is CI state, $D$ is deployment
state, and $E$ is the set of persisted evidence records. In the current implementation,
most of this state is represented through GitHub and linked execution systems.

GitHub is not treated merely as a ticket tracker. It is the human-readable and
machine-readable blackboard for collaboration. The design follows the blackboard tradition,
where independent knowledge sources contribute to a shared problem state
\citep{erman1980hearsay,corkill1991blackboard}, but adds software-delivery-specific
versioning, authority receipts, CI evidence, and explicit state transitions. Statechart ideas
also motivate treating workflow phases and invalidation as first-class state
\citep{harel1987statecharts}.

\subsection{The Six-Part Loop Contract}

An agent loop is a tuple
\[
L_i = \langle T_i, K_i, A_i, V_i, S_i, W_i \rangle,
\]
with the following elements:

\begin{description}[leftmargin=2.4cm,style=nextline]
  \item[Trigger $T_i$] A predicate over $G_t$ that determines eligibility.
  \item[Context $K_i$] A projection of authoritative state that must be reconstructed on
  every run.
  \item[Authority $A_i$] The allowed transition relation, including explicit negative
  permissions.
  \item[Verifier $V_i$] A mapping from an attempted transition and resulting state to
  independently inspectable evidence.
  \item[Stop $S_i$] Conditions that prohibit action or require return to a human.
  \item[Write-back $W_i$] A keyed, reconcilable update that persists state, evidence, and the
  reason for completion or refusal.
\end{description}

Conceptually, a loop invocation executes:

\begin{lstlisting}
if not T_i(G): return NO_OP
if not acquire_lease(item, phase): return NO_OP
c = K_i(G)
if S_i(c): return W_i(STOP, reason)
a = choose_action(c) where a is permitted by A_i
G_prime = execute(a)
e = V_i(a, G_prime)
if not acceptable(e): return W_i(STOP, e)
return W_i(ADVANCE, e)
\end{lstlisting}

The agent may use an LLM to interpret context or choose an action. Eligibility, permission,
version binding, and evidence requirements are evaluated against durable state rather than
the model's explanation. Their enforcement strength varies by implementation: some are
restricted by tool or credential boundaries, some are checked by structured validators, and
some remain procedural contracts that require audit. \Cref{tab:enforcement} makes these
distinctions explicit.

\subsection{Version-Bound Authorization}

A human approval is represented as a consumable receipt
\[
r = \langle actor, action, target, version, scope, issuedAt, consumed \rangle.
\]
Here, $\mathrm{trusted}$ checks membership in the configured approving-actor set;
$\mathrm{target}$ resolves the work item and environment named by the transition;
$\mathrm{allowed}$ checks that the requested action lies within the receipt's declared scope;
and $\mathrm{fresh}$ checks expiry and transition-specific invalidation predicates.
For a consequential transition at state $G$, the receipt is valid only if
\[
\begin{split}
\mathrm{valid}(r,G) \Leftrightarrow {}&
\mathrm{trusted}(r.actor) \land
r.target = \mathrm{target}(G) \\
&\land r.version = H(G) \land
\mathrm{allowed}(r.action,r.scope) \\
&\land \neg r.consumed \land
\mathrm{fresh}(r,G).
\end{split}
\]

If the head changes from $h$ to $h'$, then $r(h)$ is invalid even when the pull request,
branch name, or conversational intent appears unchanged. The same principle applies to plan
versions and environments: approval for plan $v_n$ does not authorize $v_{n+1}$, and a test
grant does not authorize production.

\subsection{Evidence-Gated Completion}

An evidence record is
\[
e = \langle claim, artifact, version, verifier, result, timestamp \rangle.
\]
Each state transition declares a required evidence set $\mathcal{E}_{req}$. Closure is legal
only if all required claims are supported by evidence bound to the current version:
\[
\mathrm{close}(G) \Rightarrow
\forall q \in \mathcal{E}_{req},\ \exists e \in E(G):
e.claim=q \land e.version=H(G) \land e.result=\mathrm{pass}.
\]

Evidence is layered. A local test, green CI, successful build, accepted deployment request,
runtime digest, service health check, and user-visible smoke test support different claims.
They are not interchangeable. The verifier should, where practical, use a source that did
not produce the original conclusion. In this paper, \emph{source-independent verification}
means that evidence is read from a system distinct from the initiating action---for example,
runtime identity after a deployment request. It does not necessarily imply a separate human,
organization, or model instance.

\subsection{Coordination Safety Properties}

The operational model specifies five intended safety properties that distinguish a Loop
Engineering workflow from a best-effort agent conversation:

\begin{description}[leftmargin=2.65cm,style=nextline]
  \item[Version safety] A consequential transition can advance only when its authorization,
  review, CI, and required evidence refer to the current head.
  \item[Environment separation] Authority for a lower-consequence environment never implies
  authority for a higher-consequence environment.
  \item[Evidence freshness] Evidence for $h$ cannot prove a claim about $h'$ when $h' \neq h$,
  even if both revisions share a branch or pull request.
  \item[Single-writer intent] A loop may write only when it owns the current unexpired lease
  for the work item and phase.
  \item[No implicit escalation] Missing context, ambiguous intent, or unavailable verification
  cannot expand authority; it must produce a stop result.
\end{description}

For a transition $a$ from $G_t$ to $G_{t+1}$, the core version property is
\[
\mathrm{advance}(a,G_t) \Rightarrow
\mathrm{version}(a)=H(G_t)
\land \mathrm{version}(r_a)=H(G_t)
\land \forall e \in E_a:\mathrm{version}(e)=H(G_t).
\]
The production gate adds an application-specific stronger invariant: no autonomous loop may
merge the production path or deploy the production environment.

The current lease implementation is a cooperative protocol over durable records rather than a
linearizable distributed lock. It reduces duplicate writers and supports reconciliation, but a
concurrent race remains possible between observation and write-back. Consequently,
single-writer intent is a design property to monitor, not a proven mutual-exclusion guarantee.

\subsection{Recoverability Under At-Least-Once Execution}

Schedulers and external APIs are assumed to provide at-least-once rather than exactly-once
execution. External effects and GitHub updates are not atomic. Each loop therefore:

\begin{enumerate}
  \item reconstructs current state instead of trusting session memory;
  \item acquires a phase-specific lease;
  \item probes for an existing side effect before writing;
  \item keys durable records by work item, phase, and exact version;
  \item keys write-back so a retry can recognize an existing logical effect; and
  \item reconciles live Git, CI, and runtime state after interruption.
\end{enumerate}

This makes a resumed run a reconciliation problem rather than a request to ``remember where
we were.'' It provides a recovery strategy, not a proof that every external side effect is
exactly-once. Agent Workflow Memory studies reusable action routines
\citep{wang2024awm}; Loop Engineering complements that direction by placing resumable
workflow truth outside model memory.

\section{Production Realization}

\subsection{System Layers}

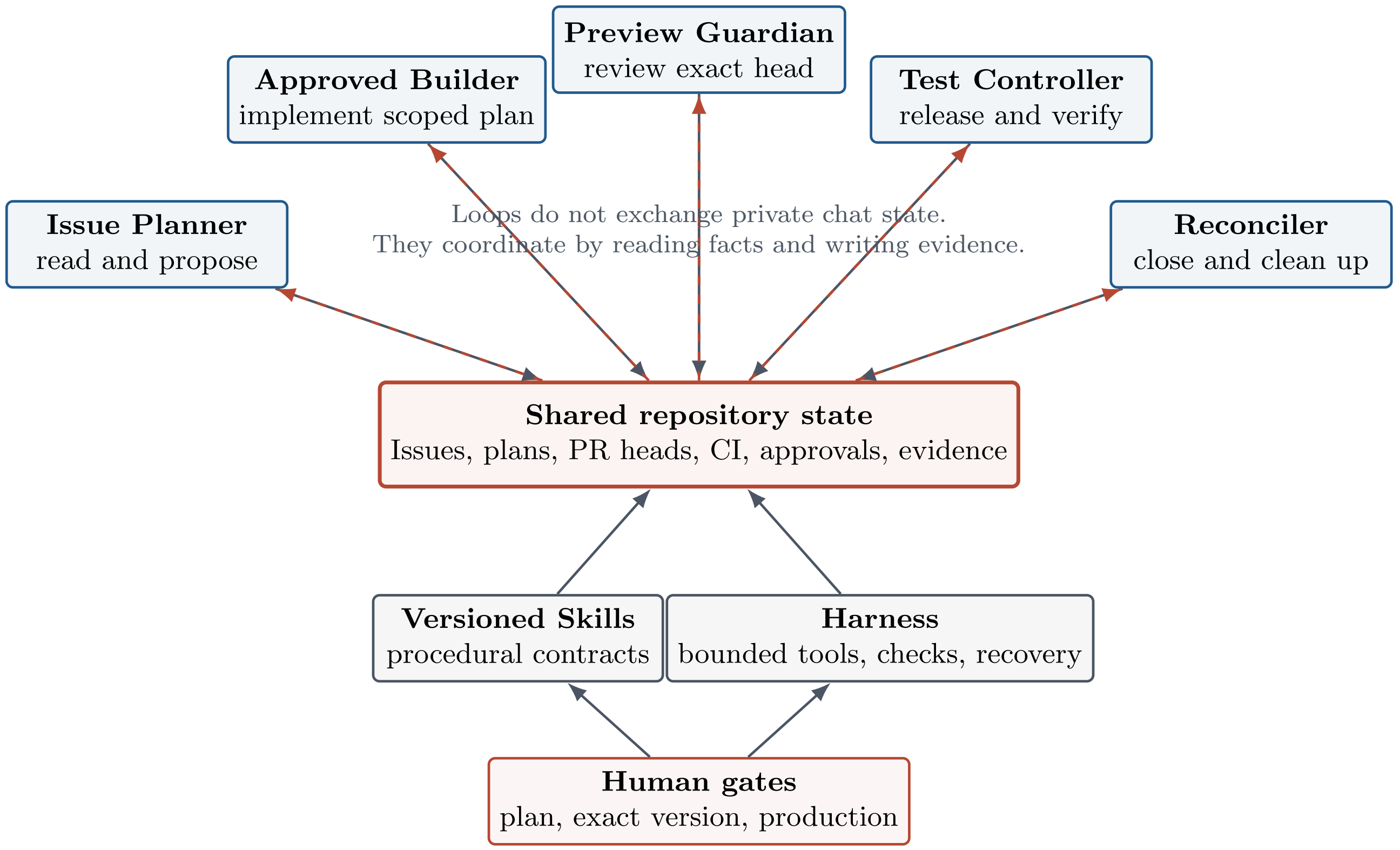
\begin{figure}[t]
  \centering
  \resizebox{\textwidth}{!}{\input{figures/architecture.tex}}
  \caption{Five asynchronous loops coordinate through shared repository state. Skills encode
  procedural contracts, the Harness routes bounded capabilities, and humans authorize
  high-consequence transitions. Solid arrows represent state reads or authorized actions;
  dashed arrows represent persisted evidence.}
  \label{fig:architecture}
\end{figure}

\Cref{fig:architecture} shows the production architecture. Four layers have distinct
responsibilities:

\begin{description}[leftmargin=2.0cm,style=nextline]
  \item[GitHub] Stores shared task state, exact revisions, decisions, and evidence.
  \item[Loops] Decide when a responsibility is eligible to run and when it must stop.
  \item[Skills] Encode versioned procedural contracts for a class of work.
  \item[Harness] Provides bounded tools and environment access, mediates structured checks,
  and collects evidence.
\end{description}

\subsection{Five Responsibility Loops}

The deployment currently uses five loops, chosen because they create meaningful authority
boundaries rather than because five roles are intrinsically desirable.

\begin{table}[t]
\centering
\small
\caption{Production loops and their intentionally missing authority.}
\label{tab:loops}
\begin{tabular}{>{\raggedright\arraybackslash}p{0.20\textwidth}
                >{\raggedright\arraybackslash}p{0.37\textwidth}
                >{\raggedright\arraybackslash}p{0.33\textwidth}}
\toprule
\textbf{Loop} & \textbf{Responsibility} & \textbf{Explicit non-authority} \\
\midrule
Issue Planner & Investigate eligible issues and publish one versioned, scoped plan
& Cannot write implementation code or approve its own plan \\
Approved Builder & Implement the exact approved plan in an isolated worktree; test and open
paired delivery paths & Cannot expand scope, reuse stale approval, or merge release branches \\
Preview Guardian & Review the live pull-request head against plan, diff, tests, and CI
& Cannot silently rewrite unrelated or unowned branches \\
Test Release Controller & Consume exact-SHA test approval, merge the approved preview head,
follow matching build, deploy, and verify runtime & Cannot deploy production or treat request
acceptance as completion \\
Master Reconciler & Reconcile evidence and local state after a human-controlled production
path completes & Cannot merge the master path or authorize production \\
\bottomrule
\end{tabular}
\end{table}

The loops do not send summaries directly to each other. The Planner writes the plan and phase
record. The Builder independently reloads the issue, approved plan version, repository, and
current base. The Guardian reads the actual diff, head, checks, and prior findings. The Test
Controller reads exact approval and obtains deployment evidence. The Reconciler reads live
merge and runtime state. The handoff is shared state, not conversational trust.

\subsection{Skills as Procedural Contracts}

\begin{minipage}{\textwidth}
In this system, a Skill is not a persona prompt. It is a version-controlled operating
contract that answers:

\begin{enumerate}
  \item When does this contract apply?
  \item Which sources must be read before acting?
  \item Which tools may be used?
  \item What may and may not be changed?
  \item Which evidence proves success?
  \item Which conditions require stopping?
\end{enumerate}
\end{minipage}

Separate Skills govern repository collaboration, testing, deployment, read-only operations,
logs, and data investigation. This separation reduces the capability set loaded for any one
task and prevents convenient but unsafe responsibility blending. A deployment Skill should
not improvise database writes; an operations Skill should not merge code; a data Skill should
not restart services. The abstraction draws on Design by Contract \citep{meyer1992dbc}, while
recent evidence on repository-level agent instructions suggests that persistent guidance can
measurably affect coding-agent execution \citep{lulla2026agentsmd}.

Skills are also where operational experience accumulates. A deployment endpoint once
accepted a request before the target runtime changed, so runtime identity and health became
mandatory evidence. Code once changed after approval, so all consequential grants became
version-bound. Interrupted runs motivated durable markers, probe-before-write checks, and
idempotency keys. These requirements are testable contract properties. Whether they are
preventive controls or auditable obligations depends on the mechanism summarized in
\cref{tab:enforcement}.

\begin{table}[H]
\centering
\small
\caption{Sanitized contract for the Test Release Controller. Product identifiers, endpoints,
credentials, and implementation commands are omitted.}
\label{tab:skill-contract}
\begin{tabular}{>{\raggedright\arraybackslash}p{0.19\textwidth}
                >{\raggedright\arraybackslash}p{0.74\textwidth}}
\toprule
\textbf{Element} & \textbf{Contract} \\
\midrule
Trigger & Current-head review ledger has no open finding; no active test-release lease exists \\
Context & Issue, approved plan, live pull-request head, current target, required checks, and
environment state are reloaded from authoritative sources \\
Authority & Merge one approved Preview revision and invoke the test deployment entry point \\
Non-authority & No production deployment, production-data write, force push, or interpretation
of request acceptance as deployment completion \\
Verifier & Match the build to the approved revision; observe runtime identity, health, bounded
logs, and a user-visible smoke result \\
Stop & Head or target drift, stale or ambiguous approval, missing checks, conflicting lease,
or runtime mismatch \\
Write-back & Persist consumed receipt, exact merge revision, evidence claims, or a sanitized
stop reason keyed by pull request and version \\
\bottomrule
\end{tabular}
\end{table}

\subsection{Harness as Responsibility Router}

The Harness combines:

\begin{itemize}
  \item repository and GitHub access;
  \item isolated worktree management;
  \item test, browser, and CI observation;
  \item deployment entry points;
  \item read-only runtime, log, and data inspection;
  \item trusted-actor validation;
  \item leases, exact-version locks, and idempotency keys; and
  \item structured evidence write-back.
\end{itemize}

The same underlying model behaves differently in a chat window and in this Harness because
the latter changes the action space, not merely the prompt. This follows least-privilege and
complete-mediation principles \citep{saltzer1975protection}: each loop receives the minimum
authority intended for its phase, and consequential accesses are checked against current
state. The realization is not a single policy engine. It combines capability restriction,
structured validation, cooperative protocol, and human control.

\begin{table}[t]
\centering
\small
\caption{Enforcement matrix for the current realization. ``Structured'' means machine-checked
by loop logic; it does not imply a formally verified or tamper-proof mechanism.}
\label{tab:enforcement}
\begin{tabular}{>{\raggedright\arraybackslash}p{0.22\textwidth}
                >{\raggedright\arraybackslash}p{0.37\textwidth}
                >{\raggedright\arraybackslash}p{0.33\textwidth}}
\toprule
\textbf{Control} & \textbf{Primary mechanism} & \textbf{Strength and residual risk} \\
\midrule
Exact-version approval & Trusted-actor lookup, exact marker grammar, head comparison, and
consumed-receipt record & Structured check; a race remains between observation and write-back \\
Scope and prohibited actions & Skill contract plus phase-specific tool and environment exposure
& Mixed control; not every prohibition is enforced below the agent layer \\
CI and runtime evidence & Exact-revision workflow lookup plus runtime identity, health, log,
and smoke probes & Machine-observed; external telemetry may be unavailable or delayed \\
Single-writer intent & Durable lease marker, re-fetch, ownership check, and expiry
& Cooperative protocol; not a linearizable distributed lock \\
Production boundary & No autonomous production transition; master merge and production release
remain human-controlled & Strong workflow boundary; human-account compromise is out of scope \\
\bottomrule
\end{tabular}
\end{table}

\subsection{Human Gates and Progressive Authority}

Human involvement is placed at high-consequence transitions:

\begin{itemize}
  \item \textbf{Plan gate:} authorize one exact plan version.
  \item \textbf{Test gate:} authorize one exact commit for the test environment.
  \item \textbf{Production gate:} retain human control over master merge and production
  release.
\end{itemize}

Database migrations, production data writes, payment or permission logic, traffic changes,
workflow cancellation, force pushes, and destructive cleanup remain outside autonomous
authority. These are deliberate responsibility boundaries, not missing features. Authority
can expand only when the system has stronger verifiers and sufficient evidence for the
additional risk.

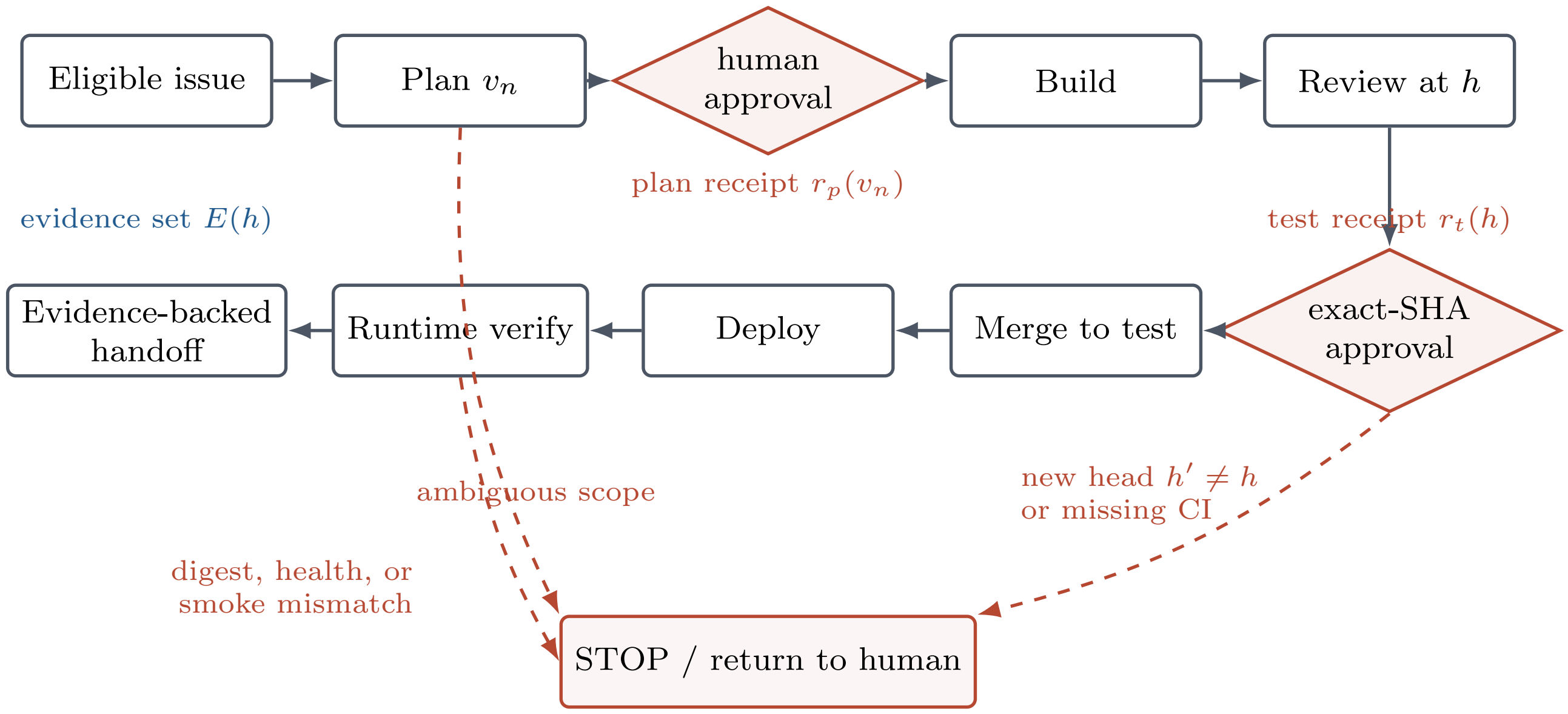
\begin{figure}[t]
  \centering
  \resizebox{\textwidth}{!}{\input{figures/state-machine.tex}}
  \caption{A simplified state machine. Human approval is a version-bound receipt. Ambiguity,
  head drift, missing checks, or runtime mismatch moves the workflow to a stop state. A new
  head invalidates prior test approval even if the pull request remains the same.}
  \label{fig:state-machine}
\end{figure}

\Cref{fig:state-machine} illustrates the main progression and invalidation paths.

\section{Early Production Evaluation}

\subsection{Questions and Method}

The evaluation asks four feasibility questions:

\begin{description}[leftmargin=0.9cm]
  \item[RQ1] Can asynchronous loops coordinate without private conversational handoffs?
  \item[RQ2] Do version-bound receipts and evidence gates stop stale progression?
  \item[RQ3] Which recovery mechanisms are present, and what recovery evidence is observed?
  \item[RQ4] Which consequential decisions still require humans?
\end{description}

We audited a cross-sectional internal snapshot collected on July 23, 2026, after migration of
the five loops. We enumerated repository-wide issue and pull-request comments containing the
canonical marker prefix, classified each record by its declared type, deduplicated by immutable
comment identifier, and counted unique containing work items. The 50 included records were
created from 2026-07-20T10:22:10Z through 2026-07-22T17:13:09Z. Mutable status records could
be updated until snapshot extraction; the latest included update was
2026-07-23T03:25:44Z. All timestamps are UTC.

The unit of observation is therefore a durable comment record, not a scheduler invocation.
Plan, phase/lease, review-ledger, test-lock, and test-evidence types contribute to the reported
count. Pull-request pair markers stored in PR bodies were inspected during case reconstruction
but are not included in the 50 comment records. For all five test-lock candidates, we inspected
associated revision history, checks, merges, approval comments, deployment records, and
available runtime verification.

Raw repository records are proprietary and may include implementation details. The paper
therefore reports aggregate counts and sanitized event sequences only. No customer content,
personal data, production database rows, prompts, secrets, endpoints, or source code were
included. The artifact contains both aggregate counts and an event-level sanitized trace for
the detailed case. Version aliases preserve equality and drift relationships without exposing
repository identifiers or commit hashes.

\begin{table}[t]
\centering
\small
\caption{Evidence strength by research question. The final column deliberately separates an
observed trace from an untested system-wide property.}
\label{tab:rq-results}
\begin{tabular}{>{\raggedright\arraybackslash}p{0.08\textwidth}
                >{\raggedright\arraybackslash}p{0.41\textwidth}
                >{\raggedright\arraybackslash}p{0.42\textwidth}}
\toprule
\textbf{RQ} & \textbf{Observed evidence} & \textbf{Supported conclusion} \\
\midrule
RQ1 & Three complete test-release paths advanced through durable records
& Feasibility in this deployment; no comparative superiority claim \\
RQ2 & One exact-head drift trace rejected prior approval and stale CI
& One observed fail-closed episode; no estimated failure rate \\
RQ3 & Durable markers, leases, probe-before-write logic, and one withheld activation
& Recovery mechanisms are implemented; recovery correctness remains untested \\
RQ4 & Exact plan/test receipts and human-only production transitions
& Descriptive authority boundary for this deployment \\
\bottomrule
\end{tabular}
\end{table}

\subsection{Artifact Audit}

\begin{table}[t]
\centering
\small
\caption{Durable coordination artifacts in the audit snapshot. Unique-item counts are not
additive because one work item can contain several marker types.}
\label{tab:audit}
\begin{tabular}{lrr}
\toprule
\textbf{Artifact} & \textbf{Marker instances} & \textbf{Unique items} \\
\midrule
Plan record & 14 & 12 \\
Phase or lease record & 17 & 17 \\
Version-bound review ledger & 7 & 7 \\
Exact-version test lock & 5 & 5 \\
Test evidence handoff & 7 & 7 \\
\midrule
All marker-bearing records & 50 & 26 \\
\bottomrule
\end{tabular}
\end{table}

The 50 records across 26 issues and pull requests show that coordination state was persisted
outside agent sessions. Evidence markers can appear at multiple handoff locations, so marker
count is not a task count.

Five pull requests acquired an exact-version test lock. Three were merged into the test path
and received runtime evidence from sources distinct from the deployment request. Two stopped
before merge. This is not reported as a success rate: tasks were heterogeneous and not sampled
as an experiment. The relevant observation is that acquiring a lock did not force progression.

\subsection{Three Verified Release Paths}

\Cref{tab:cases} summarizes all three test releases that reached terminal runtime evidence in
the audit. Repository surface is included only to show that the loops operated on multi-file
changes rather than isolated function exercises; it is not a productivity measure.

\begin{table}[t]
\centering
\small
\caption{Sanitized verified test-release cases.}
\label{tab:cases}
\begin{tabular}{>{\raggedright\arraybackslash}p{0.08\textwidth}
                >{\raggedright\arraybackslash}p{0.23\textwidth}
                >{\raggedright\arraybackslash}p{0.18\textwidth}
                >{\raggedright\arraybackslash}p{0.36\textwidth}}
\toprule
\textbf{Case} & \textbf{Task class} & \textbf{Repository surface}
& \textbf{Terminal evidence} \\
\midrule
A & Data-correctness workflow & 18 changed files
& Matching build, runtime identity, health, and smoke checks \\
B & Public-page and SEO change & 42 changed files
& Matching build, runtime identity, health, and smoke checks \\
C & Administrative state semantics & 4 changed files
& Deterministic tests, browser checks, runtime identity, zero restarts, bounded logs, and
deployed smoke \\
\bottomrule
\end{tabular}
\end{table}

For each case, the release controller consumed approval bound to the current head, merged the
matching preview revision, followed the build corresponding to that merge, and checked the
actual runtime. An accepted deployment request alone was insufficient. This answers RQ1 at
the feasibility level: planning, implementation, review, and release responsibilities advanced
through shared state without a private inter-agent conversation.

\subsection{Detailed Case: Full Path and Head Drift}

Case C provides the clearest end-to-end trace. The Planner persisted plan version 1 at
05:17 UTC. A response at 05:59 contained the approval phrase plus additional text and did not
satisfy the exact approval grammar; an exact response at 08:03 allowed the Builder to proceed.
The Builder created paired Preview and Master paths. The Guardian wrote its review record at
11:40, the Test Controller issued a lock at 12:42, and a separate exact-SHA approval arrived
at 13:52. At 14:16, the matching build and runtime were recorded as verified: the intended
runtime identity was observed, the service had zero restarts, the public port returned a
successful response, blocker logs were absent in the bounded window, and two deployed
end-to-end checks passed. The elapsed time from persisted plan to test evidence was 8 hours
59 minutes, including human waiting intervals. We did not measure active human labor.

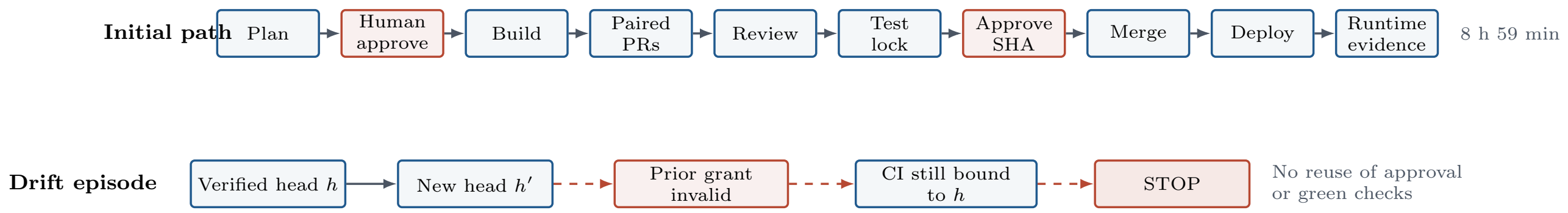
\begin{figure}[t]
  \centering
  \resizebox{\textwidth}{!}{\input{figures/evaluation-timeline.tex}}
  \caption{Top: the persisted events in Case C's first test path. Bottom: a later head-drift
  episode in which the prior version gate and CI evidence were not reused. The diagram is
  event-relative; only the end-to-end duration is shown.}
  \label{fig:timeline}
\end{figure}

The production path later exposed a more important safety episode. At 14:59, review produced
a new head $h'$ after prior test evidence had been recorded for $h$. The old test lock,
approval, and evidence were invalidated. A replacement Preview path was created at 15:01.
When the Guardian inspected it at 15:56, the synthetic merge and required CI still referred
to an earlier target relationship. The Guardian returned a stop result instead of reusing
green checks associated with the prior state. The later human-controlled merge is not counted
as automated recovery evidence. \Cref{fig:timeline} visualizes the two traces.

This episode addresses RQ2 as one observed trace. The workflow did not need to infer whether
the newer revision was ``close enough.'' Exact version mismatch made prior authority invalid.
It does not establish a population-level stale-evidence rejection rate.

\subsection{Recovery and Human Work}

Durable markers and live-state reconstruction provide the mechanism for loops to resume from
GitHub, Git, CI, and runtime facts instead of an earlier agent transcript. Duplicate planning,
pull-request creation, and release effects are checked through markers, leases, and
probe-before-write logic. A separate configuration audit detected that one migrated
automation's operational ownership did not match its declared responsibility; activation was
withheld until corrected.

These observations answer only the mechanism portion of RQ3. The snapshot does not contain a
controlled interruption or fault-injection trial that establishes recovery correctness,
duplicate-effect rate, or convergence time. Those outcomes remain open evaluation work.

For RQ4, humans remained responsible for:

\begin{itemize}
  \item choosing and clarifying product intent;
  \item approving a specific plan version;
  \item approving an exact test revision;
  \item deciding whether production consequences are acceptable; and
  \item resolving cases where evidence cannot distinguish product ambiguity from failure.
\end{itemize}

Observed human work centered on authorizing consequential state transitions rather than
issuing every implementation command. The current deployment does not automate master merge
or production release, and the audit did not quantify active human time or cognitive load.

\section{Related Work}

\subsection{Software Engineering Agents}

Repository-level benchmarks and agents establish the importance of realistic code context,
tool interfaces, and executable verification
\citep{jimenez2024swebench,yang2024sweagent,wang2025openhands}.
Agentless cautions that complexity is not itself capability \citep{xia2024agentless}, while
SWE-Lancer adds economically grounded tasks \citep{miserendino2025swelancer}. AI Agents That
Matter argues for evaluation beyond accuracy, including cost, reproducibility, and downstream
utility \citep{kapoor2024agents}. Loop Engineering does not propose a new patch-generation
policy. It contributes the surrounding operational protocol for determining when and under
what evidence a patch may progress.

\subsection{Conversational and Protocol-Based Multi-Agent Systems}

ChatDev, MetaGPT, and AutoGen organize specialized agents through communication, roles, or
standard operating procedures \citep{qian2024chatdev,hong2024metagpt,wu2024autogen}. MAST
identifies system-design, inter-agent misalignment, and verification failures
\citep{cemri2025mast}. CooperBench reports a coordination penalty when coding agents must
resolve conflicting work \citep{khatua2026cooperbench}.

SEMAP is especially close in spirit: it introduces explicit behavioral contracts,
structured messaging, lifecycle control, and verification for multi-agent systems
\citep{mao2025semap}. AgentForge likewise treats execution-grounded verification and shared
memory as first-class \citep{kumar2026agentforge}.

Two 2026 systems are still closer to the present work. Agentic Agile-V combines lifecycle
discipline, structured contracts, risk-adaptive acceptance, evidence bundles, independent
verification, and human approval \citep{koch2026agenticagilev}. CAID coordinates asynchronous
engineering agents through centralized delegation, isolated Git worktrees, structured
integration, and executable verification \citep{geng2026caid}. These systems make it
insufficient to claim novelty merely from contracts, evidence, role separation, or
asynchronous execution.

\begin{table}[t]
\centering
\small
\caption{Positioning against the closest recent approaches. The distinctions identify each
paper's primary contribution rather than claiming that an implementation cannot be extended.}
\label{tab:related-comparison}
\begin{tabular}{>{\raggedright\arraybackslash}p{0.15\textwidth}
                >{\raggedright\arraybackslash}p{0.31\textwidth}
                >{\raggedright\arraybackslash}p{0.45\textwidth}}
\toprule
\textbf{Approach} & \textbf{Primary overlap} & \textbf{Distinction of Loop Engineering} \\
\midrule
Agentic Agile-V & Structured contracts, risk-adaptive acceptance, evidence bundles, and human
approval & Repository-state runtime protocol with consumable exact-revision receipts,
asynchronous recovery records, and deployed-runtime evidence \\
CAID & Asynchronous agents, isolated Git worktrees, structured integration, and executable
tests & Lifecycle responsibilities and authority gates across planning, review, test release,
runtime verification, and reconciliation \\
SEMAP & Behavioral contracts, lifecycle control, structured coordination, and verification
& Repository records instead of a message protocol, plus version- and environment-bound
authority \\
AgentForge & Shared state and mandatory execution-grounded verification
& Operational progression around a patch, including human receipts, deployment, runtime
identity, and stop evidence \\
\bottomrule
\end{tabular}
\end{table}

Loop Engineering's narrower contribution is the combination of repository-state-centric
coordination, exact-revision and environment-bound authority receipts, resumable operational
records, and evidence that extends from review and CI through deployed runtime state. The
present study complements rather than supersedes the controlled evaluations in CAID and
SEMAP.

\subsection{Human-in-the-Loop and Workflow Memory}

HULA integrates human refinement into planning and coding within an industrial issue tracker
\citep{takerngsaksiri2025hula}. Loop Engineering similarly treats the existing collaboration
system as the human-agent surface, but concentrates human effort at explicit consequential
gates and persists machine-readable transition receipts. AWM learns reusable workflows for
future agent trajectories \citep{wang2024awm}; Skills instead encode reviewed procedural
contracts, while the shared state makes recovery independent of model memory.

\subsection{Blackboards, State Machines, and Least Authority}

The blackboard architecture provides the closest classical coordination analogy:
independent specialists observe and modify shared problem state
\citep{erman1980hearsay,corkill1991blackboard}. Statecharts motivate explicit transitions
and exceptional paths \citep{harel1987statecharts}. Loop Engineering adapts these ideas to
versioned repository work, where commit identity, CI, and deployment evidence are central.
Its authority model also follows fail-safe defaults, complete mediation, and least privilege
\citep{saltzer1975protection}.

\section{Discussion}

\subsection{Why Non-Chatting Loops Matter}

The claim is not that agents should never communicate. Discussion is useful when negotiation
itself is the task. The claim is narrower: a private conversation is a poor source of truth
for operational state that must survive restarts, be inspected by humans, and authorize
consequential actions.

Shared state changes the semantics of coordination:

\begin{itemize}
  \item A handoff is a persisted state transition, not a message saying ``your turn.''
  \item A review is a claim bound to a revision, not confidence in the builder.
  \item Approval is a scoped, consumable receipt, not sentiment extracted from conversation.
  \item Completion is a set of evidence claims, not the executing agent's final answer.
  \item Memory is reconstructed from records, not assumed to remain in context.
\end{itemize}

This also makes the number of agents secondary. If two responsibilities share the same
trigger, context, authority, verifier, and stop conditions, splitting them into two personas
adds little. A new loop is justified by a real boundary: different authority, independent
source evidence, asynchronous waiting, or a distinct recovery lifecycle.

\subsection{Skills as Organizational Memory}

Skills turn individual operational experience into versioned organizational memory. Their
value is not the volume of instructions but the precision of boundaries and tests. A useful
Skill changes after an incident: what was once tacit becomes an explicit precondition,
invalidation rule, evidence requirement, or stop condition.

This suggests a maturation path for agentic engineering:

\begin{enumerate}
  \item prompts describe desired behavior;
  \item Skills define repeatable procedures and prohibited actions;
  \item the Harness bounds the available action surface;
  \item evidence tests the contract;
  \item incident review updates the contract and its verifier.
\end{enumerate}

The durable asset is therefore not a transcript with a capable model. It is the evolving set
of contracts, tools, and evidence rules that make different model instances more bounded,
auditable, and recoverable.

\subsection{Toward Better Evaluation}

Patch success alone does not measure the outer loop. Future evaluations should include:

\begin{itemize}
  \item \textbf{Unauthorized transition rate:} attempted or completed actions outside authority.
  \item \textbf{Stale-evidence rejection:} whether version drift invalidates prior claims.
  \item \textbf{Evidence completeness:} required claims supported by source-distinct artifacts.
  \item \textbf{Recovery correctness:} convergence after interruption, retry, or partial effects.
  \item \textbf{Duplicate-effect rate:} repeated writes, deployments, or notifications.
  \item \textbf{Human gate load:} number, latency, and consequence of required interventions.
  \item \textbf{Time to verified outcome:} not time to generated patch.
  \item \textbf{Audit reconstruction:} whether an independent reviewer can explain each
  consequential transition from durable records.
\end{itemize}

Controlled comparison between conversational multi-agent systems, one long-running agent,
and shared-state loops would be valuable. Fault injection should vary head changes, delayed
CI, duplicate scheduler invocations, ambiguous approvals, deployment acceptance without
runtime change, and partial external failure.

\section{Limitations and Threats to Validity}

\paragraph{Single setting.}
The deployment concerns one commercial product, one primary system designer, one repository
collaboration platform, and a short post-migration window. Results may not generalize to
larger teams, regulated environments, monorepos, or non-GitHub workflows.

\paragraph{No causal baseline.}
We did not randomize tasks or compare equivalent work under chat-based, single-agent, and
Loop Engineering conditions. We therefore make no causal claim about productivity, quality,
cost, or revenue.

\paragraph{Selection and measurement.}
The observed tasks are production work, not a representative sample. Durable markers measure
recorded state transitions, not every scheduler invocation or no-op. Changed-file counts are
descriptive and are not a productivity measure.

\paragraph{Enforcement validity.}
The implementation combines hard capability boundaries, structured checks, cooperative
protocols, and natural-language contracts. It is not a formally verified reference monitor.
In particular, GitHub comments and external side effects are not atomic, and leases are not
linearizable. The enforcement matrix exposes these distinctions, but the present audit does
not measure bypass attempts or adversarial behavior.

\paragraph{Human-agent attribution.}
Automation writes through an authenticated human-owned GitHub identity. Public authorship
metadata cannot cleanly distinguish human text from automation-generated text. Our analysis
therefore uses artifact type and transition sequence, not author identity, as the unit of
evidence.

\paragraph{Confidentiality.}
Raw repository, deployment, operational, and customer records cannot be released. This limits
external replication. We provide aggregate counts, operational contracts, a sanitized
event-level trace, and buildable manuscript source.

\paragraph{Model evolution.}
The workflow has used multiple coding-agent generations. Model changes, prompt changes,
Skill changes, and infrastructure changes are confounded. The method is intended to be
model-agnostic, but the present deployment cannot isolate model effects.

\paragraph{Observer bias.}
The primary system designer also performed the artifact extraction and analysis. Future work
should include independent coding of events, additional organizations, pre-registered
measures, fault injection, and adversarial audits.

\section{Conclusion}

The central engineering problem for production coding agents is not how to keep them running
indefinitely. It is how to make goals, state, authority, evidence, and stop conditions form a
recoverable closed loop.

Loop Engineering treats the repository collaboration system as durable shared memory and a
state machine. Specialized loops wake on explicit predicates, reconstruct authoritative
context, operate under least authority, verify outcomes with version-bound evidence, and
write the result back for the next human or agent. Skills preserve operational contracts; the
Harness bounds capabilities and mediates evidence checks; humans retain control over
consequential transitions.

The MewDesign deployment provides early evidence that this architecture can coordinate
multiple asynchronous responsibilities without private conversational handoffs and progress
real changes through a test environment with runtime evidence. One drift trace shows that the
workflow can stop rather than reuse stale approval and CI. The evidence is limited, but the
systems lesson is concrete: stronger models do not remove the need for engineering control.
They increase the value of making that control explicit, auditable, and reusable.

\section*{Artifact and Ethics Statement}

The accompanying source package contains the LaTeX manuscript, BibTeX references, TikZ
figures, aggregate evidence, and a sanitized event-level trace. It contains no proprietary
code, raw private-repository records, customer content, personal data, production database
rows, secrets, or internal endpoints. The audit analyzed engineering process artifacts
created during ordinary product development; it did not study end users or customer behavior.
Commercial context is provided only to establish that the workflow operated in a live product
rather than a demonstration, and is not used as an outcome measure.

\section*{Competing Interests}

All authors are affiliated with MewDesign, the commercial product studied in this paper. This
relationship constitutes a professional and potentially financial interest. The paper reports
process evidence rather than product efficacy, revenue impact, or comparative commercial
performance.

\section*{Generative AI Use Disclosure}

Generative AI tools assisted literature discovery, manuscript structuring, language editing,
LaTeX drafting, and consistency checking. The listed human authors remain responsible for the
research design, evidence interpretation, source verification, claims, and final manuscript.

\section*{Author Contributions}

\textbf{Yupeng Pang:} Conceptualization, methodology, software, investigation, system
architecture, production implementation, project administration, and iterative practice
optimization.
\textbf{Meiling Chen:} Validation and overall Loop system testing.
\textbf{Yang Yang:} Software, validation, Test Loop optimization, and implementation
refinement.
\textbf{Wei Fu:} Writing---original draft and writing---review and editing.

\bibliographystyle{plainnat}
\bibliography{references}

\end{document}

%% file: figures/architecture.tex
\begin{tikzpicture}[
  font=\small,
  loop/.style={draw=loopblue, thick, rounded corners=2pt, fill=loopblue!6,
    minimum width=3.1cm, minimum height=0.95cm, align=center},
  layer/.style={draw=loopgray, thick, rounded corners=2pt, fill=loopgray!5,
    minimum width=3.2cm, minimum height=0.95cm, align=center},
  state/.style={draw=loopred, very thick, rounded corners=2pt, fill=loopred!6,
    minimum width=5.0cm, minimum height=1.15cm, align=center},
  arrow/.style={-{Latex[length=2.2mm]}, thick, draw=loopgray},
  evidence/.style={-{Latex[length=2.2mm]}, thick, dashed, draw=loopred}
]
  \node[state] (github) at (0,0) {\textbf{Shared repository state}\\
  Issues, plans, PR heads, CI, approvals, evidence};

  \node[loop] (planner) at (-6.1,2.1) {\textbf{Issue Planner}\\read and propose};
  \node[loop] (builder) at (-3.45,3.7) {\textbf{Approved Builder}\\implement scoped plan};
  \node[loop] (guardian) at (0,4.25) {\textbf{Preview Guardian}\\review exact head};
  \node[loop] (test) at (3.45,3.7) {\textbf{Test Controller}\\release and verify};
  \node[loop] (reconcile) at (6.1,2.1) {\textbf{Reconciler}\\close and clean up};

  \node[layer] (skills) at (-2.0,-2.25) {\textbf{Versioned Skills}\\procedural contracts};
  \node[layer] (harness) at (2.0,-2.25) {\textbf{Harness}\\bounded tools, checks, recovery};
  \node[layer, fill=loopred!5, draw=loopred] (human) at (0,-4.05)
    {\textbf{Human gates}\\plan, exact version, production};

  \foreach \n in {planner,builder,guardian,test,reconcile} {
    \draw[arrow] (\n) -- (github);
    \draw[evidence] (github) -- (\n);
  }
  \draw[arrow] (skills) -- (github);
  \draw[arrow] (harness) -- (github);
  \draw[arrow] (human) -- (skills);
  \draw[arrow] (human) -- (harness);

  \node[align=center, text=loopgray, font=\footnotesize] at (0,2.25)
    {Loops do not exchange private chat state.\\They coordinate by reading facts and writing evidence.};
\end{tikzpicture}

%% file: figures/state-machine.tex
\begin{tikzpicture}[
  font=\footnotesize,
  state/.style={draw=loopgray, thick, rounded corners=2pt, fill=white,
    minimum width=2.15cm, minimum height=0.78cm, align=center},
  gate/.style={draw=loopred, thick, diamond, aspect=2.1, fill=loopred!7,
    inner sep=1.5pt, align=center},
  arrow/.style={-{Latex[length=2mm]}, thick, draw=loopgray},
  invalid/.style={-{Latex[length=2mm]}, thick, dashed, draw=loopred}
]
  \node[state] (backlog) at (0,0) {Eligible issue};
  \node[state] (plan) at (2.7,0) {Plan $v_n$};
  \node[gate] (plangate) at (5.35,0) {human\\approval};
  \node[state] (build) at (8.0,0) {Build};
  \node[state] (review) at (10.7,0) {Review at $h$};

  \node[gate] (testgate) at (10.7,-2.15) {exact-SHA\\approval};
  \node[state] (merge) at (8.0,-2.15) {Merge to test};
  \node[state] (deploy) at (5.35,-2.15) {Deploy};
  \node[state] (verify) at (2.7,-2.15) {Runtime verify};
  \node[state] (done) at (0,-2.15) {Evidence-backed\\handoff};

  \draw[arrow] (backlog) -- (plan);
  \draw[arrow] (plan) -- (plangate);
  \draw[arrow] (plangate) -- (build);
  \draw[arrow] (build) -- (review);
  \draw[arrow] (review) -- (testgate);
  \draw[arrow] (testgate) -- (merge);
  \draw[arrow] (merge) -- (deploy);
  \draw[arrow] (deploy) -- (verify);
  \draw[arrow] (verify) -- (done);

  \node[font=\scriptsize, text=loopred] at (5.35,-0.9) {plan receipt $r_p(v_n)$};
  \node[font=\scriptsize, text=loopred] at (10.7,-1.2) {test receipt $r_t(h)$};
  \node[font=\scriptsize, text=loopblue] at (0,-1.2) {evidence set $E(h)$};

  \node[state, draw=loopred, fill=loopred!5] (stop) at (5.35,-5.0)
    {STOP / return to human};
  \draw[invalid] (plan.south) to[bend right=14] (stop.north west);
  \draw[invalid] (testgate.south) to[bend left=12] (stop.north east);
  \draw[invalid] (verify.south) to[bend right=10] (stop.west);
  \node[font=\scriptsize, text=loopred, align=right] at (3.35,-3.55)
    {ambiguous scope};
  \node[font=\scriptsize, text=loopred, align=left] at (8.55,-3.55)
    {new head $h' \neq h$\\or missing CI};
  \node[font=\scriptsize, text=loopred, align=right] at (1.25,-4.35)
    {digest, health, or\\smoke mismatch};
\end{tikzpicture}

%% file: figures/evaluation-timeline.tex
\begin{tikzpicture}[
  font=\scriptsize,
  event/.style={rectangle, rounded corners=1.5pt, draw=loopblue, thick, fill=loopblue!5,
    minimum width=1.35cm, minimum height=0.62cm, align=center, inner sep=1.5pt},
  gate/.style={rectangle, rounded corners=1.5pt, draw=loopred, thick, fill=loopred!8,
    minimum width=1.35cm, minimum height=0.62cm, align=center, inner sep=1.5pt},
  stop/.style={rectangle, rounded corners=1.5pt, draw=loopred, thick, fill=loopred!12,
    minimum width=1.5cm, minimum height=0.62cm, align=center, inner sep=1.5pt},
  arrow/.style={-{Latex[length=1.8mm]}, thick, draw=loopgray},
  stale/.style={-{Latex[length=1.8mm]}, thick, dashed, draw=loopred}
]
  \node[event] (p1) at (0,0) {Plan};
  \node[gate] (p2) at (1.65,0) {Human\\approve};
  \node[event] (p3) at (3.3,0) {Build};
  \node[event] (p4) at (4.95,0) {Paired\\PRs};
  \node[event] (p5) at (6.6,0) {Review};
  \node[event] (p6) at (8.25,0) {Test\\lock};
  \node[gate] (p7) at (9.9,0) {Approve\\SHA};
  \node[event] (p8) at (11.55,0) {Merge};
  \node[event] (p9) at (13.2,0) {Deploy};
  \node[event] (p10) at (14.85,0) {Runtime\\evidence};
  \foreach \a/\b in {p1/p2,p2/p3,p3/p4,p4/p5,p5/p6,p6/p7,p7/p8,p8/p9,p9/p10} {
    \draw[arrow] (\a) -- (\b);
  }
  \node[anchor=east, font=\footnotesize\bfseries] at (-0.35,0) {Initial path};
  \node[anchor=west, text=loopgray] at (15.7,0) {8 h 59 min};

  \node[event, minimum width=2.05cm] (d1) at (0,-2.0) {Verified head $h$};
  \node[event, minimum width=2.05cm] (d2) at (2.75,-2.0) {New head $h'$};
  \node[gate, minimum width=2.3cm] (d3) at (5.75,-2.0) {Prior grant\\invalid};
  \node[event, minimum width=2.4cm] (d4) at (9.0,-2.0) {CI still bound\\to $h$};
  \node[stop, minimum width=2.05cm] (d5) at (12.0,-2.0) {STOP};
  \draw[arrow] (d1) -- (d2);
  \draw[stale] (d2) -- (d3);
  \draw[stale] (d3) -- (d4);
  \draw[stale] (d4) -- (d5);
  \node[anchor=east, font=\footnotesize\bfseries] at (-1.35,-2.0) {Drift episode};
  \node[anchor=west, text=loopgray, align=left] at (13.2,-2.0)
    {No reuse of approval\\or green checks};
\end{tikzpicture}